\documentclass[prd,twocolumn,amssymb,altaffilletter,runningaddress,superscriptaddress,showpacs,showkeys,epsfig]{revtex4}
\usepackage[dvips]{graphicx}
\usepackage[english]{babel}
\newcommand{\be}{\begin{equation}}
\newcommand{\ee}{\end{equation}}
\newcommand{\ben}{\begin{eqnarray}}
\newcommand{\een}{\end{eqnarray}}

\begin{document}

\title{Evolution of a Kerr-Newman black hole in a dark energy universe}
\author{Jos\'e A. Jim\'enez Madrid}
\email{madrid@imaff.cfmac.csic.es}\affiliation{Instituto de Matem\'aticas y F\'\i sica Fundamental
Consejo Superior de Investigaciones Cient\'\i ficas,
Serrano 121, 28006 Madrid, Spain}
\affiliation{Instituto de Astrof\'\i sica de Andaluc\'\i a, Consejo Superior de Investigaciones Cient\'\i ficas, Camino Bajo de Hu\'etor 50, 18008 Granada, Spain}
\author{Pedro F. Gonz\'alez-D\'\i az}\email{p.gonzalezdiaz@imaff.csic.es}\affiliation{Instituto de Matem\'aticas y F\'\i sica Fundamental
Consejo Superior de Investigaciones Cient\'\i ficas,
Serrano 121, 28006 Madrid, Spain}
\date{\today}

\begin{abstract}
    {This paper deals with the study of the accretion of dark energy with 
equation of state $p=w\rho$ onto Kerr-Newman black holes. We have 
obtained that when $w>-1$ the mass and specific angular 
momentum increase, and that whereas the specific angular 
momentum increases up to a given plateau, the mass
grows up unboundedly. On the regime where the dominant
energy condition is violated our model predicts a steady decreasing
of mass and angular momentum of black holes as phantom energy is being
accreted. Masses and angular momenta of all black holes tend to
zero when one approaches the big rip. The results that cosmic
censorship is violated and that the black hole
size increases beyond the universe size itself are discussed in terms
of considering the used models as approximations to a more general 
descriptions where the metric is time-dependent.}

\end{abstract}

\maketitle

\section{Introduction}

Several astronomical  and cosmological observations,  ranging from
 observations of distant supernovae Ia\cite{Riess:1998cb} to
the cosmic microwave background anisotropy\cite{Spergel:2003cb}, 
indicate  that
the universe is currently undergoing an accelerating stage. It is assumed that
this acceleration is due to some unknown stuff usually dubbed dark
energy, with a positive energy density $\rho>0$ and
with negative pressure $p<-(1/3)\rho.$
There are several candidate models for describing the dark energy, being the cosmological constant, $\Lambda$, by far
the simplest and most popular candidate\cite{Lambda}. Other
interesting models are based on considering a perfect fluid with given
equation of state like in quintessence\cite{quintessence}, K-essence\cite{Armendariz-Picon:1999rj} or 
generalized Chaplygin gas models\cite{chaplygin,RBOD,MMG,chaplygin2,MBL-PVM}.
Note  that
there are also other candidates for dark energy based on brane-world 
models\cite{brane} and modified 4-dimensional Einstein-Hilbert
actions\cite{EHmodified}, where a late time acceleration of the
universe may be achieved, too.

One of the peculiar properties of the resulting cosmological models
is the possibility of occurrence of  a cosmic doomsday, also dubbed big
rip\cite{phantom1,caldwell,phantom3,SSD,Li,phantom2}.
The big rip appears in models where dark energy 
particularizes as the so-called phantom energy for which
$p+\rho<0$. In these models the scale factor
blows up in a finite time because its cosmic acceleration is even
larger than that  induced by a positive cosmological constant.
In these models every component of the universe  goes beyond the 
horizon of all other universe components in
finite cosmic time. It should be noted, that the condition $p+\rho<0$
is not enough for the occurrence of a big rip\cite{Bouhmadi-Lopez:2004me}.
In recent papers\cite{Babichev:2004yx,Babichev:2005py}, it has been shown
that the mass of a Schwarzschild black hole decreases with accretion
of phantom energy, in such a way that the black hole disappears at the time 
of  the
big rip. Therefore, it is interesting to study how dark energy is accreted 
by more general black holes, that is to say, black holes bearing charge and
angular momentum. The interest of this study is enhanced by the eventual competition or joint contribution that may arise between the dark energy
accretion process and super-radiance which tends to decrease the
rotational (or charge) energy  of the hole, so lowering
its spin (or charge), such as one would expect phantom energy induced 
as well. For this reason, in the present paper we shall investigate how distinct forms 
of dark energy can be accreted onto Kerr-Newman black holes. We in fact obtain
that Kerr-Newman black holes progressively increase their mass and angular
momentum as a result from dark energy accretion when the equation of
state allows $p+\rho>0$. That increase of mass and angular momentum
is either unbounded or tends to a given plateau, depending on the dark
energy model being considered. If $p+\rho<0$ then both
the mass and the angular momentum of black hole rapidly decrease until 
disappearing at the big rip, or tend to constant values in the absence of
a future singularity. It is seen that the latter process prevails over both the
Hawking evaporation process and spin super-radiance.
Our quantitative results appear to indicate, on the other hand, 
that whereas phantom energy does not violate cosmic 
censorship conjecture\cite{Penrose}, dark energy with $w>-1$ does.

The paper can be outlined as follows. In the next section, we will 
generalize the solution obtained by  Babichev, Dokuchaev and Eroshenko\cite{Babichev:2004yx,Babichev:2005py}  to
the case of  dark energy accretion onto a charged, rotating black hole, and
present the general equations for the rate of mass and momentum. In the next section we apply such a formalism to quintessence and K-essence
cosmological fields, so as to the generalized Chaplygin gas model, analyzing
the corresponding evolution of the black hole. 
In section \ref{approximated} we discuss the results that cosmic censorship 
is violated and that the black hole size grows up unboundedly beyond
the universe size in terms of considering the used models as
approximations to a more general description where the metric is not static.
Finally, we  briefly summarize and discuss our
results in section \ref{conclusion}

\section{General accretion formalism  for  Kerr-Newman black holes}
\label{sec:formalism}
In this section we shall follow the accretion formalism, first 
considered by Babichev, Dokuchaev and Eroshenko\cite{Babichev:2004yx,Babichev:2005py}, generalizing it to the case in which
the black hole has an angular momentum and charge. First of all, we notice
that, even though we shall use a static Kerr-Newman metric, the time 
evolution induced by accretion will be taken into account by the time 
dependence of the scale factor entering the integrated conservation laws 
and the rate equations for mass and angular momentum.

The procedure is based on integrating the conservation laws
for energy-momentum tensor and its projection onto the 
four-velocity, using as general definition of energy-momentum 
tensor a perfect fluid where the properties of
the dark energy and those of the black hole metric are both contained.
By combining the results from these integrations with 
assumed rate equations for black hole mass, angular momentum and
specific angular momentum, we can derive final rate equations for
these quantities in terms of the dark pressure, $p$, and dark
energy density, $\rho.$ Now, since the conservation of dark energy and its
state equation $p=w\rho$ lead to a unique relation between
$p$ and $\rho$ with the scale factor $R(t)$, our final rate equation
will only depend on $R(t).$

Using a static metric nevertheless restrict in 
principle ourselves to deal with small
accretion rates as the mixed component of  the energy-momentum tensor 
used in this case to derive the
metric is zero. So, at first sight, this procedure 
becomes an approximate scheme whose description can only be
valid for a short initial time. However, the use of a 
non-static metric for which that 
energy-momentum component
is no longer vanishing does not generally amount to different results 
asymptotically, which is the physically
relevant situation we have to consider. This question will be dealt with
in more detail in Sec.~\ref{approximated}
Throughout this
paper we shall use natural units so that $G=c=1$. 
Let us then consider the stationary and axisymmetric 
Kerr-Newman space-time. The metric in 
this case can be given by

\begin{eqnarray}
\mathrm{d} s^2&=&\left(1+\frac{Q^2-2Mr}{r^2+a^2\cos^2 \theta}\right) \mathrm{d} t^2 \nonumber \\ 
&+&\frac{2a\left(2Mr-Q^2\right)\sin^2 \theta}{r^2+a^2\cos^2 \theta}\mathrm{d} t \mathrm{d}\phi\nonumber \\
&-&\frac{r^2+a^2\cos^2\theta}{r^2+a^2+Q^2-2Mr}\mathrm{d} r^2\nonumber \\ 
&-&\left(r^2+a^2\cos^2\theta\right)\mathrm{d}\theta^2 \nonumber \\
&-&\left[\left(r^2+a^2+\frac{2Mra^2\sin^2\theta-Q^2a^2\sin^2\theta}{r^2+a^2\cos^2\theta}\right)\nonumber \right.\\
&\times& \left.\sin^2\theta\mathrm{d}\phi^2\right],
\end{eqnarray}
where $M$ is the mass, $Q$ is the electric charge, $a=J/M$ is the specific angular 
momentum of black
hole, with $J$ the total angular momentum, $r$ is the radial coordinate, and $\theta$ and $\phi$ are the
angular spherical coordinates. We model the dark energy in the black hole
by the test perfect fluid with a negative pressure and  an arbitrary 
equation of state $p(\rho)$, with the energy-momentum tensor

\begin{equation} \label{eq:deftensorenergy}
T_{\mu\nu}=\left(p+\rho\right)u_\mu u_\nu -pg_{\mu\nu},
\end{equation}
where $p$ is the pressure, $\rho$ is the energy density, and $u^\mu=dx^\mu/ds$
is the 4-velocity with $u^\mu u_\mu=1.$ 
There is no loss of generality in a restricting
consideration to $T_{\mu\nu}$ of this form, as it is actually the
most general form that $T_{\mu\nu}$ can take consistent with
homogeneity and isotropy\footnote{In this situation, $p$ and
$\rho$ are space-independent.}.

Using the general expression 
for a derivative operator\cite{Wald} applied to this case, we get that
the zeroth (time) 
component of the energy-momentum conservation law $T^{\mu\nu}{}_{;\nu}=0$ 
can then generally be written as

\begin{eqnarray}   
& &\frac{\mathrm{d}}{\mathrm{d} r}\left[\left(p+\rho\right)\left(1+\frac{Q^2-2Mr}{r^2+a^2\cos^2\theta}\right)\frac{\mathrm{d} t}{\mathrm{d} s}\frac{\mathrm{d} r}{\mathrm{d} s}\right] \nonumber \\
& &+\frac{2r}{r^2+a^2\cos^2\theta}\left(p+\rho\right)\left(1+\frac{Q^2-2Mr}{r^2+a^2\cos^2\theta}\right)\frac{\mathrm{d} t}{\mathrm{d} s}\frac{\mathrm{d} r}{\mathrm{d} s}\nonumber \\
& &+\frac{\mathrm{d}}{\mathrm{d}\theta}\left[\left(p+\rho\right)\left(1+\frac{Q^2-2Mr}{r^2+a^2\cos^2\theta}\right)\frac{\mathrm{d} t}{\mathrm{d} s}\frac{\mathrm{d}\theta}{\mathrm{d} s}\right]\nonumber \\
& &+ \left[\left(\frac{\cos\theta}{\sin \theta}-\frac{2a^2\sin\theta\cos\theta}{r^2+a^2\cos^2\theta}\right)\nonumber\right. \\
& &\left.\times\left(p+\rho\right)\left(1+\frac{Q^2-2Mr}{r^2+a^2\cos^2\theta}\right)\frac{\mathrm{d} t}{\mathrm{d} s}\frac{\mathrm{d}\theta}{\mathrm{d} s}\right]=0. \label{eq:energymomentumconservation}
\end{eqnarray} 

This expression should now be integrated.
We consider two cases. First, we take $\theta$ as a constant.
The integration of Eq.~(\ref{eq:energymomentumconservation}) gives then,
\begin{eqnarray}
C_{{\rm M}}=&\frac{u}{M^2}\left(r^2+a^2\cos^2\theta\right)\left(p+\rho\right)\nonumber \\
\times&\left(1+\frac{Q^2-2Mr}{r^2+a^2\cos^2\theta}+\frac{r^2+a^2\cos^2\theta+Q^2-2Mr}{r^2+a^2+Q^2-2Mr}u^2\right)^{1/2},\label{eq:energyconservationm}
\end{eqnarray}
where $u=\mathrm{d} r/\mathrm{d} s,$ and $C_{{\rm M}}$ is an integration constant.

Another integral of motion can be derived by using the projection of the 
conservation law for energy-momentum tensor along the four-velocity, i.e. the flux equation

\begin{equation}
u_\mu T^{\mu\nu}{}_{;\nu}=0.
\end{equation}

For a perfect fluid, this equation reduces to
\begin{equation}
u^\mu\rho_{,\mu}+\left(p+\rho\right)u^\mu{}_{;\mu}=0. \label{eq:energyvelocityconservation}
\end{equation}
The integration of Eq.~(\ref{eq:energyvelocityconservation}) gives the
second integral of motion that we shall use in what follows

\begin{equation}
\frac{u}{M^2}\left(r^2+a^2\cos^2\theta\right)\exp\left[\int^\rho_{\rho_\infty}\frac{\mathrm{d}\rho^\prime}{p\left(\rho^\prime\right)+\rho^\prime}\right]=-A_{{\rm M}},\label{eq:energyfluxequationm}
\end{equation}
where $u<0$ in the case of a fluid flow directed toward the black hole, and
$A_{{\rm M}}$ is a positive dimensionless constant. 
Eq.~\ref{eq:energyfluxequationm} gives us the energy flux induced
in the accretion process. From Eqs. (\ref{eq:energyconservationm}) and (\ref{eq:energyfluxequationm}) one can easily get:

\begin{eqnarray}
& &
\hspace{-1cm}\left(1+\frac{Q^2-2Mr}{r^2+a^2\cos^2\theta}+\frac{r^2+a^2\cos^2\theta+Q^2-2Mr}{r^2+a^2+Q^2-2Mr}u^2\right)^{1/2} \nonumber \\
& &\times\left(p+\rho\right)\exp\left[-\int^\rho_{\rho_\infty} \frac{\mathrm{d}\rho^\prime}{p\left(\rho^\prime\right)+\rho^\prime}\right]=C_{2{\rm M}}, \label{eq:finalm}
\end{eqnarray}
where $C_{2{\rm M}}=-C_{{\rm M}}/A_{{\rm M}}=p(\rho_\infty)+\rho_\infty.$

The rate of change of the black hole mass due 
to accretion 
of dark energy can be derived by integrating over the surface area 
the density of momentum $T_0 {}^r$, that is\cite{Landau}  

\begin{equation} \label{eq:rateM}
\dot{M}=-\int T_0 {}^r \mathrm{d} A, 
\end{equation}
with $\mathrm{d}A=r^2\sin\theta \mathrm{d}\theta \mathrm{d}\phi,$ and $r$ constant. Using Eqs.~(\ref{eq:deftensorenergy}), (\ref{eq:energyfluxequationm}) and 
(\ref{eq:finalm}) this can be rewritten as

\begin{equation} \label{eq:ratem}
\dot{M}=\frac{4\pi A_{{\rm M}} M^3r}{J}\arctan\left(\frac{J}{Mr}\right)\left[p\left(\rho_\infty\right)+\rho_\infty\right],
\end{equation}
with $r$ and $J$ constants. It is worth noticing that
Eq.~(\ref{eq:ratem}) consistently reduces to the corresponding
rate equation for a Schwarzschild  
black hole derived by Babichev, Dokuchaev and Eroshenko 
in Refs.~\cite{Babichev:2004yx} and \cite{Babichev:2005py} when one
lets $J$ to become very small. One has the following integral 
expression that governs the evolution of the mass of the 
Kerr-Newman black hole

\begin{eqnarray} \label{eq:integraleqm}
& &\int_{M_0}^M \frac{J \mathrm{d} M}{M^3 r \arctan \left(\frac{J}{Mr}\right)}\nonumber \\
&=&4 \pi A_{{\rm M}}\int_{t_0}^t \left[p\left(\rho_\infty\right)+\rho_\infty\right] \mathrm{d} t. 
\end{eqnarray}

Now, the integration in the left-hand-side of  Eq.~(\ref{eq:integraleqm}) gives
\begin{eqnarray} 
I(M)&=&\int_{M_0}^M \frac{J \mathrm{d} M}{M^3 r \arctan \left(\frac{J}{Mr}\right)} \nonumber \\
&=&
-\frac{r}{2J}
\times\left[ \frac{1+\frac{J^2}{M^2 r^2}}{\arctan\left( \frac{J}{Mr}\right)} 
+\frac{J}{Mr \arctan^2\left(\frac{J}{Mr}\right)}\right. \nonumber \\
& + &\frac{4}{\arctan^3\left(\frac{J}{Mr}\right)}\nonumber \\
&\times&\left.\sum_{k=1}^\infty\frac{\left(2^{2k}-1\right)\arctan^{2k}\left(\frac{J}{Mr}\right)}{\pi^{2k}\left(2k-3\right)}\zeta\left(2k\right)\right]^M_{M_0}, \label{eq:defim}
\end{eqnarray} 
where $\zeta$ is the Riemann zeta function. The integration of the right-hand-side of Eq.~(\ref{eq:integraleqm}) will be performed in the next section.
We turn now to consider $r$, instead of $\theta$, as a constant with 
which the integration of Eq.~(\ref{eq:energymomentumconservation}) yields
\begin{eqnarray}
C_{{\rm a}}&= &\frac{\omega}{a}\sin\theta\left(r^2+a^2\cos^2\theta\right)\left(p+\rho\right)\nonumber \\
&\times &\left[1+\frac{Q^2-2Mr}{r^2+a^2\cos^2\theta}\right.\nonumber\\
&+&\left.\left(r^2+a^2\cos^2\theta+Q^2-2Mr\right)\omega^2\right]^{1/2},\label{eq:energyconservationa}
\end{eqnarray}
where $\omega=\mathrm{d}\theta/\mathrm{d} s,$ and $C_{{\rm a}}$ is another integration constant.

The second integral of motion for the energy flux in this case is also obtained from the 
projection of the 
energy-momentum tensor conservation law along the four-velocity; then 
the integration of Eq.~(\ref{eq:energyvelocityconservation}) gives that
second integral of motion

\begin{equation}
\frac{1}{a}\omega\sin\theta\left(r^2+a^2\cos^2\theta\right)\exp\left[\int^\rho_{\rho_\infty}\frac{\mathrm{d}\rho^\prime}{p\left(\rho^\prime\right)+\rho^\prime}\right]=-A_{{\rm a}},\label{eq:energyfluxequationa}
\end{equation}
where $\omega<0$ in the case of a fluid flow directed toward the black hole,
and  $A_{{\rm a}}$ is a positive dimensionless constant. From Eqs.~(\ref{eq:energyconservationa}) and (\ref{eq:energyfluxequationa}) one can easily get:

\begin{eqnarray}
& &\left[1+\frac{Q^2-2Mr}{r^2+a^2\cos^2\theta}\right.\nonumber\\
&+&\left.\left(r^2+a^2\cos^2\theta+Q^2-2Mr\right)\omega^2\right]^{1/2} \nonumber \\
&\times &\left(p+\rho\right)\exp\left[-\int^\rho_{\rho_\infty} \frac{\mathrm{d}\rho^\prime}{p\left(\rho^\prime\right)+\rho^\prime}\right]=C_{2{\rm a}}, \label{eq:finala}
\end{eqnarray}
where $C_{2{\rm a}}=-C_{{\rm a}}/A_{{\rm a}}=p(\rho_\infty)+\rho_\infty.$

We take the rate of change of the specific angular momentum of the 
Kerr-Newman black hole originating from  accretion of dark energy to be now 
given by\cite{Landau}

\begin{equation} \label{eq:ratea}
\dot{a}=-\int r T_0 {}^\theta \mathrm{d} A, 
\end{equation}
with $\mathrm{d}A=r^2\sin\theta \mathrm{d}\theta \mathrm{d}\phi,$ and $\theta$ constant. Using Eqs.~(\ref{eq:energyfluxequationa}) and 
(\ref{eq:finala}) this can be rewritten as
\begin{equation}
\dot{a}=\frac{2 \pi^2 A_{{\rm a}} a r^2\left[\rho_\infty+p(\rho_\infty)\right]}{\sqrt{r^2+a^2}}
\end{equation}
with $r$ constant. Therefore, one has the following integral expression that
reports about the evolution of the specific angular momentum
\begin{equation}\label{eq:integraleqa}
\int_{a_0}^a \frac{\sqrt{r^2+a^2}}{ar^2} \mathrm{d} a =2 \pi^2 A_{{\rm a}}\int_{t_0}^t \left[p\left(\rho_\infty\right)+\rho_\infty\right]\mathrm{d} t.
\end{equation}

Then, the integration in the left-hand-side of Eq.~(\ref{eq:integraleqa})
gives rise to the following expression

\begin{eqnarray}
\hspace{-0.5cm}I(a)&=&\int_{a_0}^a \frac{\sqrt{r^2+a^2}}{ar^2} \mathrm{d} a\\
\hspace{-0.5cm}&=&\frac{1}{r^2}\left\{\sqrt{a^2+r^2}-\frac{1}{r}\ln \left[\frac{2r}{a}\left(r+\sqrt{a^2+r^2}\right)\right]\right\}^a_{a_0}.\nonumber \label{eq:defia}
\end{eqnarray}

The integration of the right-hand-side of Eq.~(\ref{eq:integraleqa}) will
again be calculated in the next section for the distinct dark energy models.

Now, we study the influence of dark energy accretion in the angular 
momentum $J$. Using $J=Ma$ and Eqs.~(\ref{eq:rateM}) and (\ref{eq:ratea})
 we can obtain the rate of change of the angular momentum 
of the black hole performing the following integral

\begin{equation}
\dot{J}=-\int \left(M r T_0 {}^\theta +a T_0 {}^r\right)\mathrm{d} A, 
\end{equation}
with $\mathrm{d}A=r^2\sin\theta \mathrm{d}\theta \mathrm{d}\phi,$ and $r$ constant. So, we obtain
\begin{eqnarray}
\dot{J}=&\pi\left[p\left(\rho_\infty\right)+\rho_\infty\right]\nonumber \\
&\times\left[\frac{2J\pi A_{{\rm a}} r}{\sqrt{1+\frac{J^2}{M^2 r^2}}}+4 A_{{\rm M}} M^2 r\arctan\left(\frac{J}{Mr}\right)\right],
\end{eqnarray} 
with $M$ and $r$ constants. Therefore, one has the following integral
expression that governs the evolution of the angular momentum of a black hole

\begin{eqnarray}
\int_{J_0}^J\frac{\mathrm{d} J}{\frac{2J\pi A_{{\rm a}} r}{\sqrt{1+\frac{J^2}{M^2 r^2}}}+4 A_{{\rm M}} M^2 r \arctan \left(\frac{J}{Mr}\right)}\nonumber\\
=\pi\int_{t_0}^t \left[p\left(\rho_\infty\right)+\rho_\infty\right] \mathrm{d} t.\label{eq:integraleqj}
\end{eqnarray} 

Once again the integration of the right-hand-side of the equation will be carried out in the next section. Here, we have been however unable to perform
the integration of the left-hand side (L) in closed form. Thus, we have proceed as follows. The integral L, in Eq.~(\ref{eq:integraleqj}) can be recast in the form

\begin{equation}
L=\int_{x_0}^x\frac{\mathrm{d} x}{\cos^2 x\left(2r\pi A_{{\rm a}}\sin x+4 A_{{\rm M}} Mx\right)}, 
\end{equation}
where $0\leq x\leq \pi/2$ and $x=\arctan\left(J/Mr\right)$. It can be noticed that, since 
$0\leq\sin x\leq x$, we have
\begin{eqnarray}
L &\geq &\int_{x_0}^x \frac{\mathrm{d} x}{\left(2\pi A_{{\rm a}} r+4A_{{\rm M}} M\right)x\cos^2 x} \nonumber\\
&=&\frac{1}{2\pi A_{{\rm a}} r+4A_{{\rm M}} M}\nonumber\\
&\times&\left[\frac{\tan x}{x}+\ln x\ 
+\frac{1}{x^2}\sum_{k=2}^\infty \frac{\left(2^{2k}-1\right) x^{2k}}{\left(k-1\right)\pi^{2k}}\zeta\left(2k\right)\right]_{x_0}^x\nonumber \\
&=&\frac{1}{2\pi A_{{\rm a}} r+4A_{{\rm M}} M}\nonumber\\
&\times&\left[\frac{J}{Mr\arctan\left(\frac{J}{Mr}\right)}+\ln\arctan\left(\frac{J}{Mr}\right)\right.\nonumber\\
&+&\frac{1}{\arctan^2\left(\frac{J}{Mr}\right)}\\
&\times&\left.\sum_{k=2}^\infty\frac{\left(2^{2k}-1\right)\arctan^{2k}\left(\frac{J}{Mr}\right)}{\left(k-1\right)\pi^{2k}}\zeta\left(2k\right)\right]_{J_0}^J
\equiv I(J), \nonumber\label{eq:defij} 
\end{eqnarray}
where $\zeta$ is again the Riemann zeta function. Thus, $L\geq I(J)$ which in
turn implies that if we use $I(J)$ for studying the evolution of the Kerr-Newman
black hole and the cosmic censorship is taken to be physically preserved, then
$L$ should respect this conjecture, too. This argument entitles us to use
$I(J)$ as a suitable expression to study the evolution of $J$ during accretion
of dark energy.

\section{Cosmological models}

In order to obtain exact integrated expressions for
the right-hand-side of Eqs.~(\ref{eq:integraleqm}), (\ref{eq:integraleqa})
 and (\ref{eq:integraleqj}), we shall use in this section two different
models for dark energy, namely, quintessence and generalized
Chaplygin gas models. It can be seen  that the results obtained by using the
quintessence model are the same as those derived if one 
used the so-called K-essence model for dark energy.

\begin{center}
\textbf{A. Quintessence models}
\end{center}

Starting with the equation of state $p=w\rho$, where $w$ is assumed
constant, we can use the conservation of the energy-momentum tensor to 
get

\begin{equation}
\rho=\rho_0\left(\frac{R_0}{R}\right)^{3\left(1+w\right)},
\end{equation}
where $R\equiv R(t)$ is the scale factor, with $\rho_0$ and $R_0$ constants. Hence

\begin{eqnarray}
&\int_{t_0}^t \left[p\left(\rho_\infty\right)+\rho_\infty\right] \mathrm{d} t\nonumber \\
=&\left(1+w\right)\rho_0 R_0^{3\left(1+w\right)}\int_{t_0}^t R^{-3\left(1+w\right)} \mathrm{d} t.
\end{eqnarray}

\begin{figure}
\begin{center}
\includegraphics[width=1.0\columnwidth]{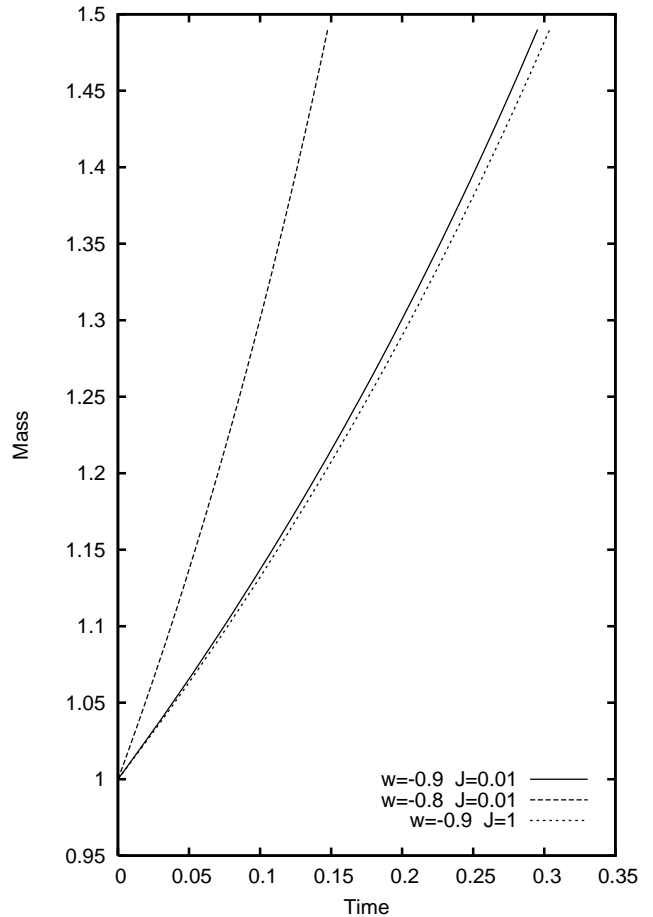}
\caption{This figure shows the behaviour of the mass of a Kerr-Newman
black hole as a function  of the cosmic time in presence of
dark energy with $w=-0.8$ and $w=-0.9$. One can also see on figure
that the larger $w$ or smaller $J$, the quicker the increase of mass.} \label{fig:m-0.9}
\end{center}
\end{figure}

\begin{figure}
\includegraphics[width=1.0\columnwidth]{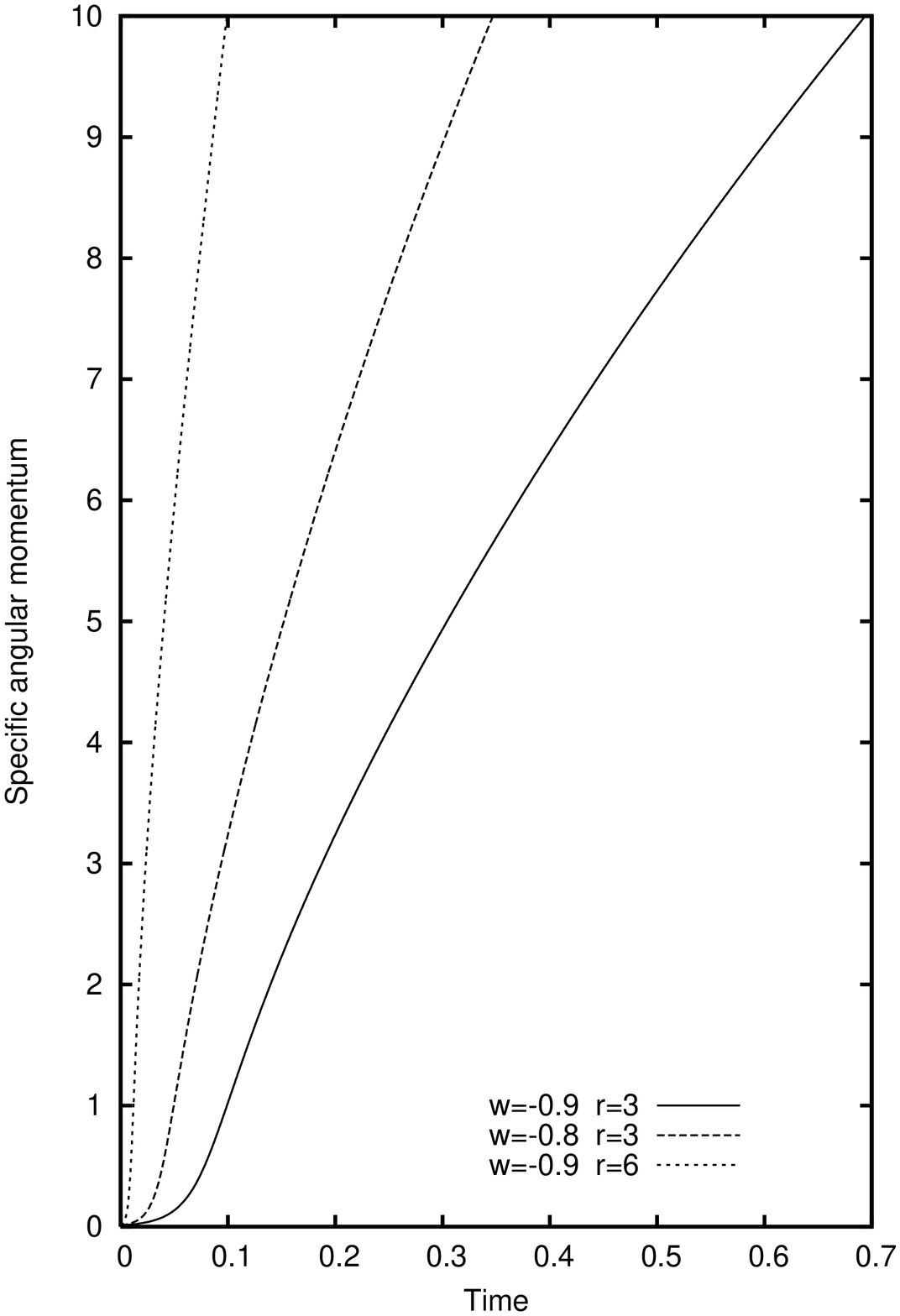}
\caption{This figure shows the behaviour of the specific angular
momentum of a Kerr-Newman
black hole as a function  of the cosmic time in presence of
dark energy with $w=-0.8$ and $w=-0.9$. One can also observe on the figure 
that the larger $w$ or $r$, the quicker the increase of specific angular momentum. In the inset we can see that $a$ tends to a constant value for
large times in all studied cases. } \label{fig:a-0.9}
\includegraphics[bb=-50 -850 50 -750, scale=0.19]{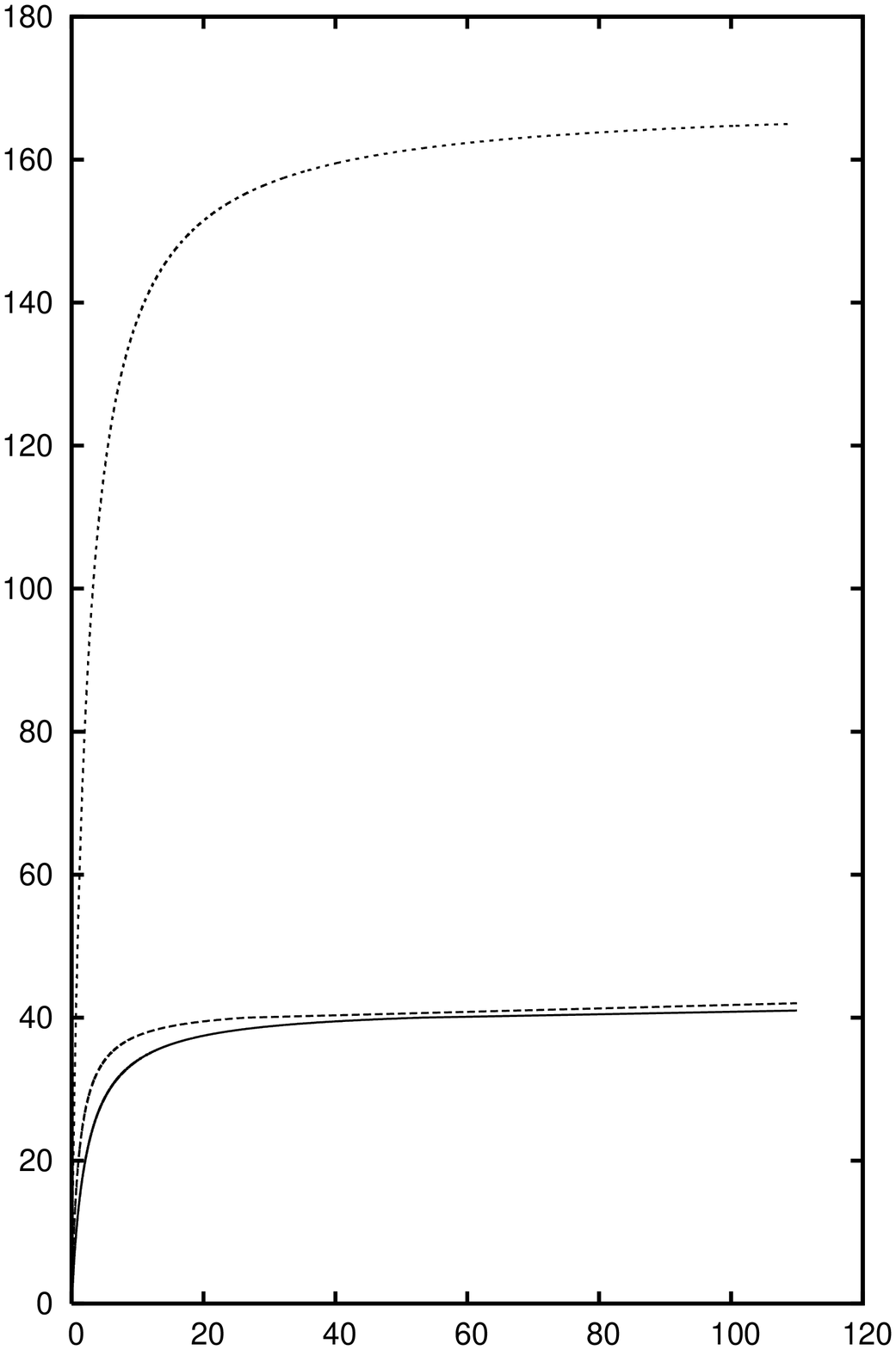}
\end{figure}

\begin{figure}
\includegraphics[width=1.0\columnwidth]{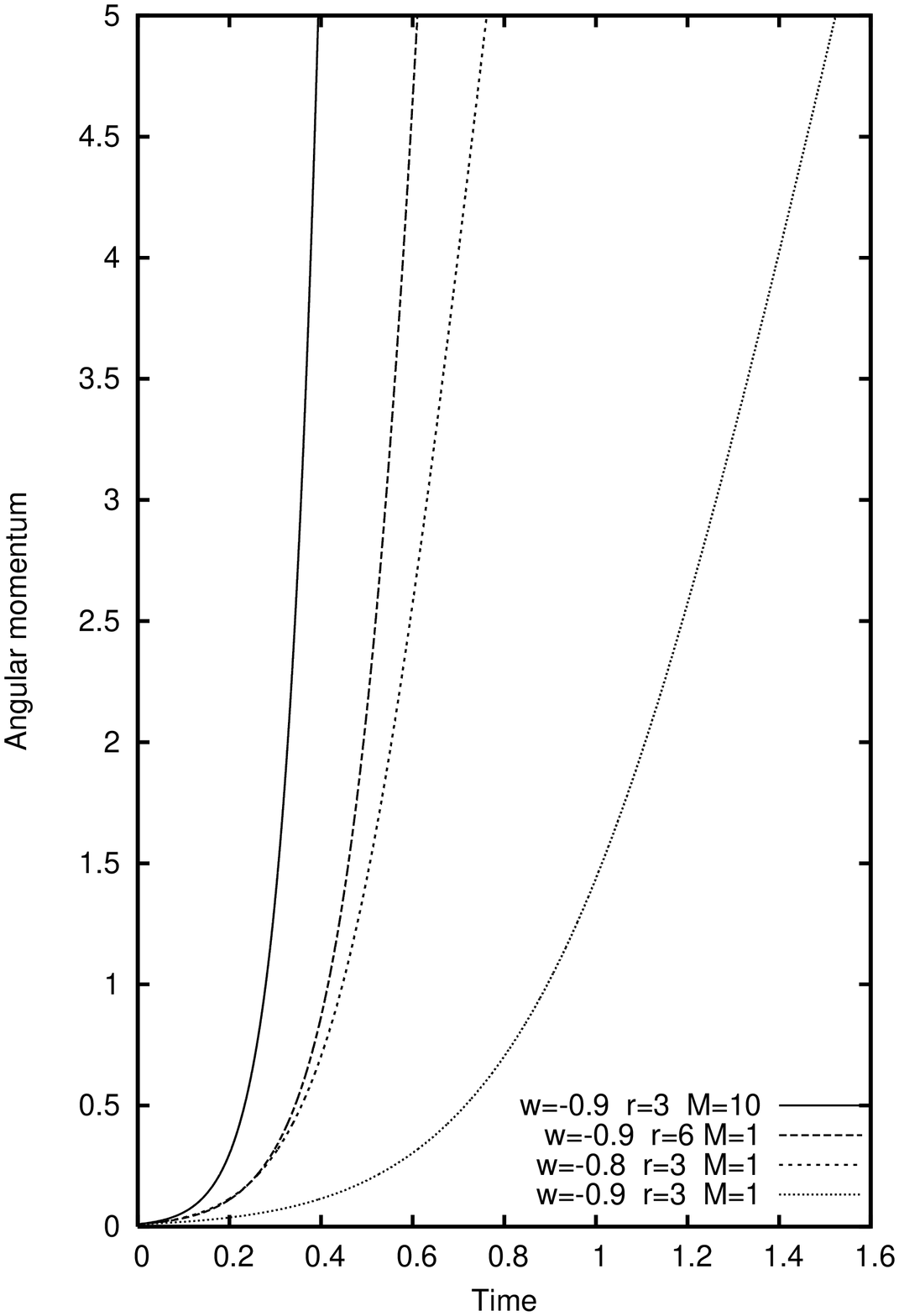}
\caption{This figure shows the behaviour of the angular
momentum of a Kerr-Newman
black hole as a function  of the cosmic time in the presence of
dark energy with $w=-0.8$ and $w=-0.9$. One can also see on the figure 
that the larger $w$, $r$ or $M$, the quicker the increase of angular momentum.
In the inset we can see that $J$ tends to a constant value for large times in
all studied cases.} \label{fig:j-0.9}
\includegraphics[bb=-350 -1200 -250 -1100, scale=0.13]{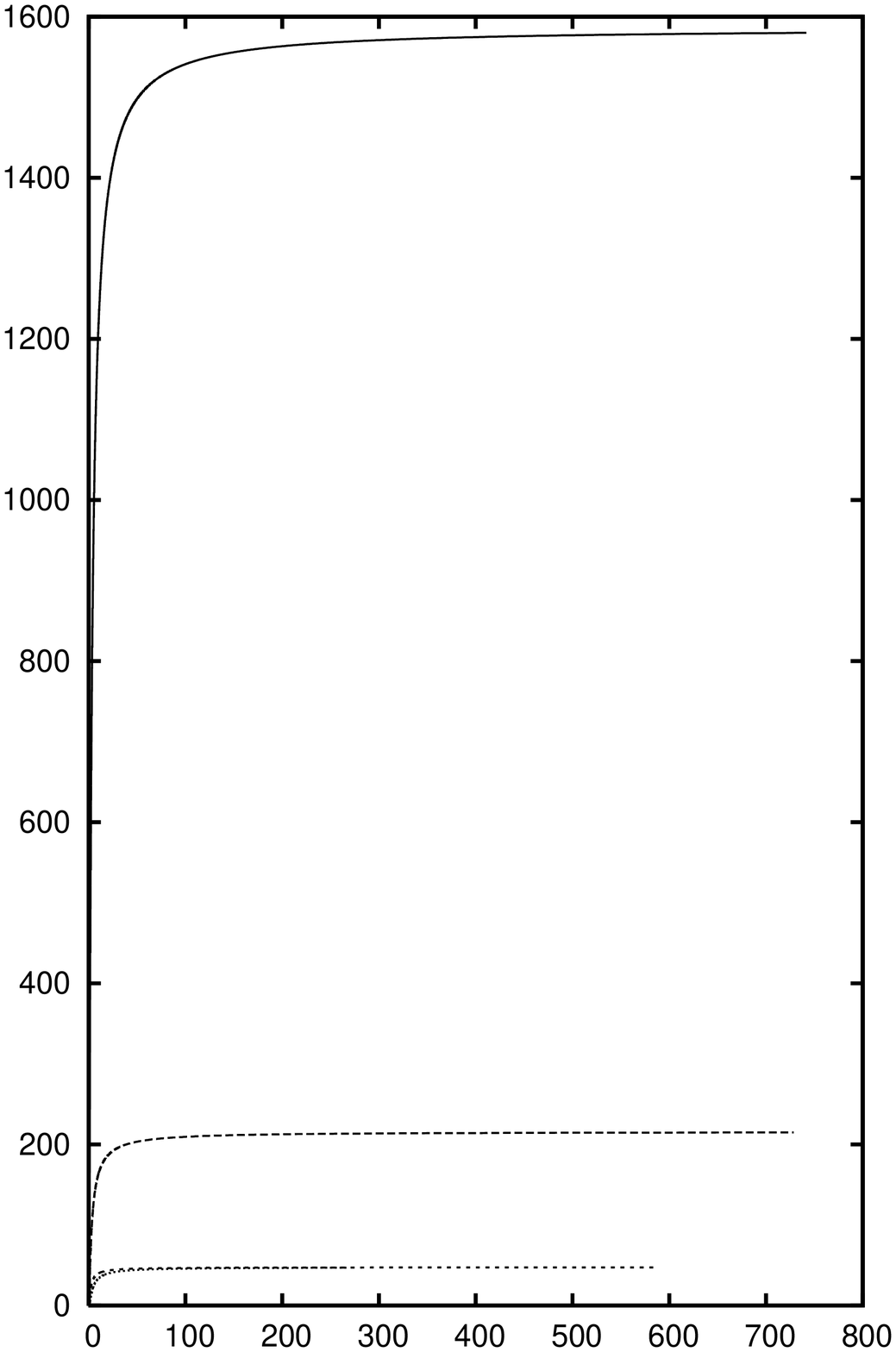}
\end{figure}

We then have for the scale factor\cite{EscalaFRW} corresponding to a general flat quintessence universe

\begin{equation}\label{eq:factorflatFRW}
R(t)=R_0\left(1+\frac{3}{2}(1+w)C^{1/2}(t-t_0)\right)^{2/[3(1+w)]},
\end{equation} 
where $C=8\pi G\rho_0/3$. Integration of the right-hand-side of 
Eqs.~(\ref{eq:integraleqm}), (\ref{eq:integraleqa})
 and (\ref{eq:integraleqj}), can then be performed using Eq.~(\ref{eq:factorflatFRW}). We respectively get    
\begin{equation}\label{eq:evolutionM}
t=t_0+\frac{I(M)}{\left(1+w\right)\left(4\pi A_{{\rm M}} \rho_0-\frac{3}{2}C^{1/2} I(M)\right)},
\end{equation}

\begin{equation}
t=t_0+\frac{I(a)}{\left(1+w\right)\left(2 \pi^2 A_{{\rm a}} \rho_0-\frac{3}{2}C^{1/2} I(a)\right)},
\end{equation}

\begin{equation}
t=t_0+\frac{I(J)}{\left(1+w\right)\left(\pi \rho_0-\frac{3}{2}C^{1/2} I(J)\right)},
\end{equation}
where $I(M), I(a) $ and $I(J)$ are defined in Eqs.~(\ref{eq:defim}), (\ref{eq:defia}) and (\ref{eq:defij}), respectively. These are three parametric
equations from which one can obtain how the mass, specific angular momentum and
angular momentum evolve in the accelerating universe. Thus, if $w>-1$ we see
that $M, a,$ and $J$ will all progressively increase with time from their initial values,
with $a$ tending to a finite constant value as $t\rightarrow\infty$, showing that
the 
increase of $M$ tends to become proportional to the increase 
of $J$; however $M$ goes to infinity in a finite time, but $J$ 
tends to a finite constant value as $t\rightarrow\infty$. Notice that
there is no contradiction between the results of 
Figs.~(\ref{fig:a-0.9}) and (\ref{fig:j-0.9}) as the plot in Fig.~(\ref{fig:j-0.9}) is obtained relative to a constant value of mass. The larger $w$ the quicker the increase
of these parameters [see Figs.~(\ref{fig:m-0.9}), (\ref{fig:a-0.9}) and (\ref{fig:j-0.9})].

\begin{figure}
\begin{center}
\includegraphics[width=1.0\columnwidth]{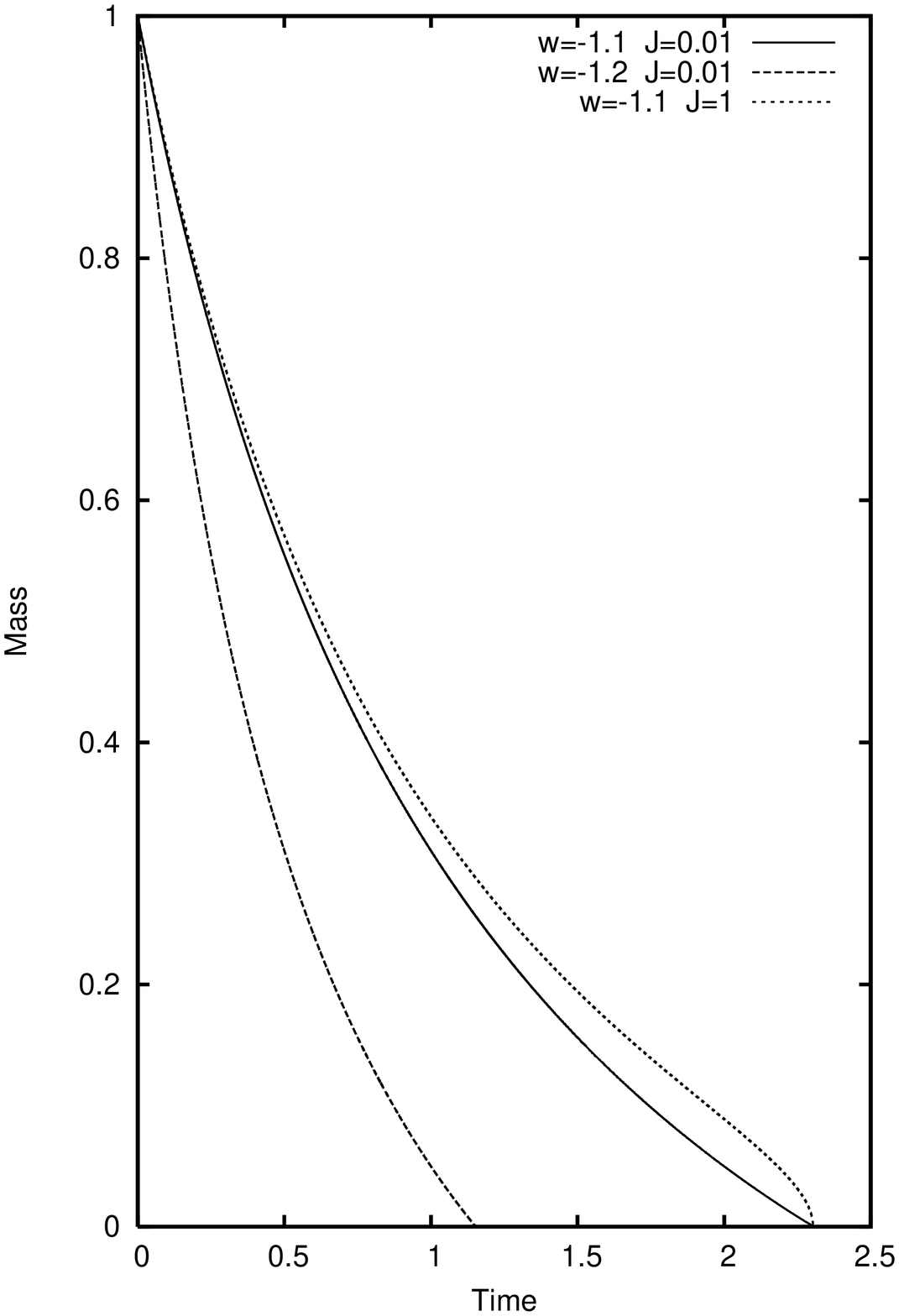}
\caption{This figure shows the behaviour of the mass of a Kerr-Newman
black hole as a function  of the cosmic time in presence of
phantom energy with $w=-1.1$ and $w=-1.2$. One can also see on the figure
that the larger $|w<-1|$ or smaller $J$, the quicker the decrease of mass.} \label{fig:m-1.1}
\end{center}
\end{figure}

\begin{figure}
\begin{center}
\includegraphics[width=1.0\columnwidth]{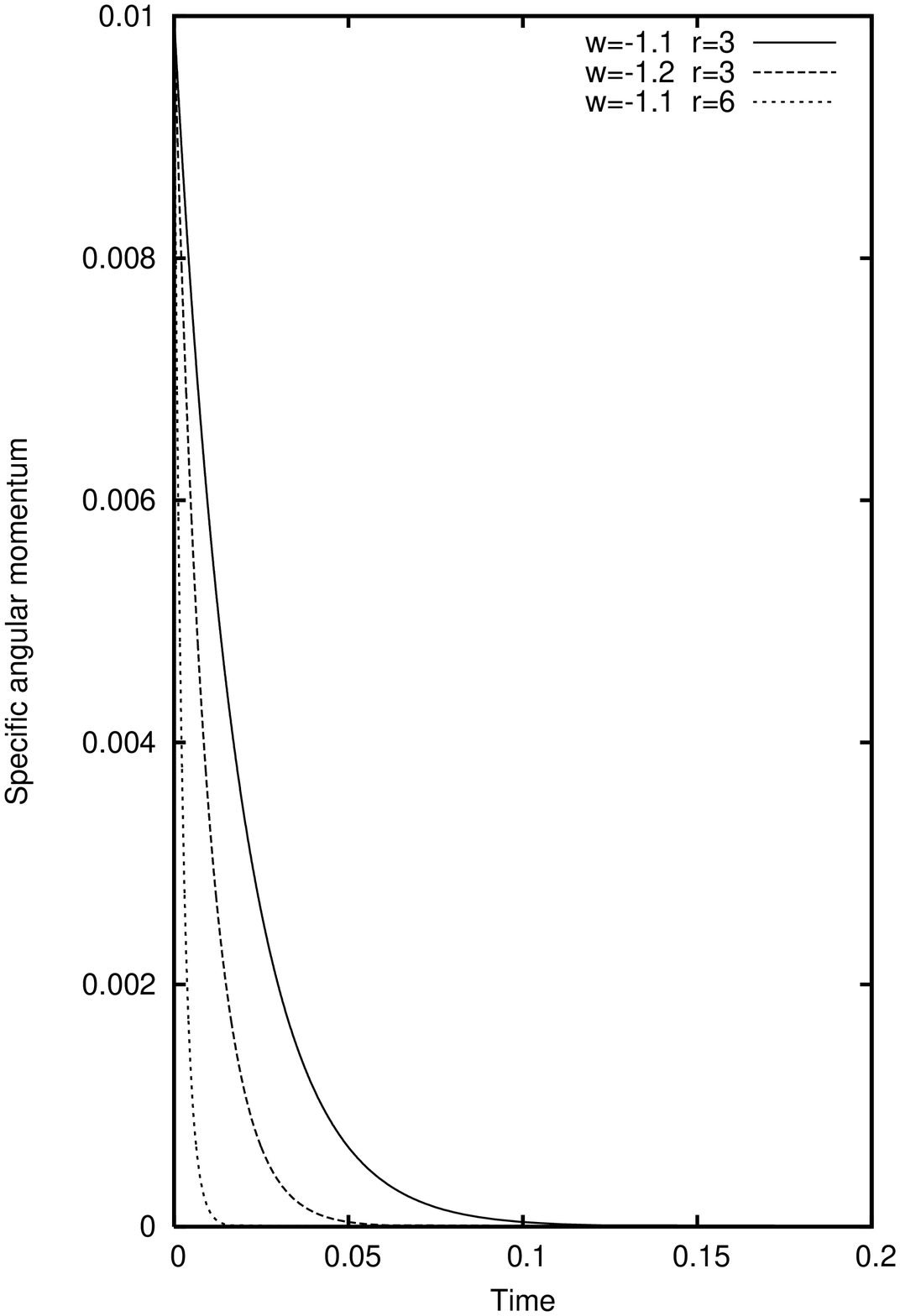}
\caption{This figure shows the behaviour of the specific angular
momentum of a Kerr-Newman
black hole as a function  of the cosmic time in the presence of
phantom energy with $w=-1.1$ and $w=-1.2$. One can also observe on the figure 
that the larger $|w<-1|$ or $r$, the quicker the decrease of specific angular momentum. } \label{fig:a-1.1}
\end{center}
\end{figure}

\begin{figure}
\begin{center}
\includegraphics[width=1.0\columnwidth]{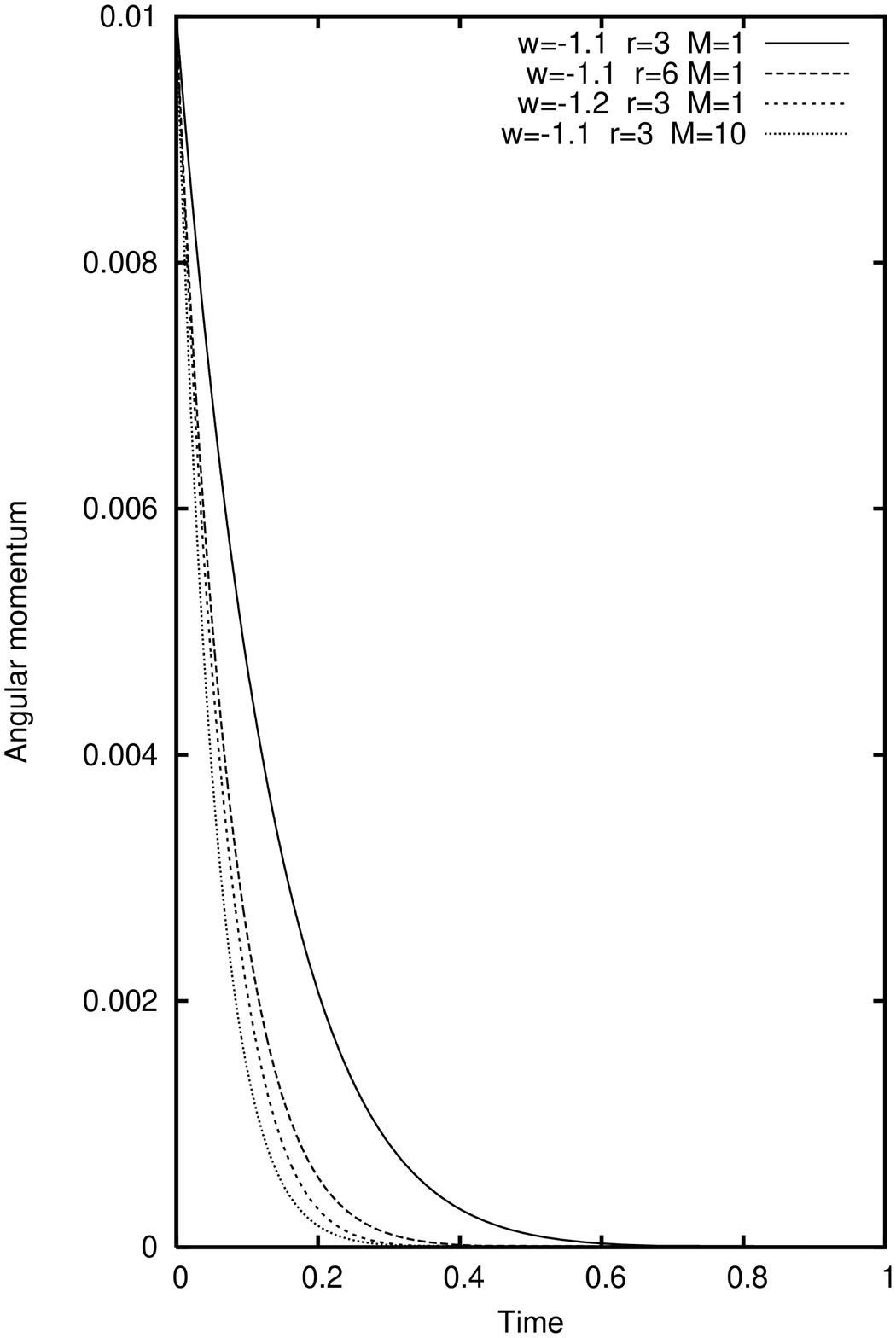}
\caption{This figure shows the behaviour of the angular
momentum of a Kerr-Newman
black hole as a function  of the cosmic time in the presence of
phantom energy with $w=-1.1$ and $w=-1.2$. One can also see on the figure 
that the larger $|w<-1|$, $r$ or $M$, the quicker the decrease of angular momentum.} \label{fig:j-1.1}
\end{center}
\end{figure}

If $w<-1$ we can observe that $M, a$, and $J$ will all progressively decrease
from their initial values, tending to zero as one approaches the big rip, where
the black holes will disappear independently of the initial values of their
mass and angular momentum [see Figs.~(\ref{fig:m-1.1}), (\ref{fig:a-1.1}) and (\ref{fig:j-1.1})]. This generalizes the result obtained by 
Babichev, Dokuchaev and Eroshenko\cite{Babichev:2004yx,Babichev:2005py}. 

In the case of a charged
black hole, the process of super-radiance of charge allows the black hole 
to emit the charge before it disappears.
It has been checked as well that the larger $|w<-1|$ the quicker is
the decrease of $M$ and $J$, and that for large $r$ the evolution
of the mass nearly matches the evolution that was derived for the 
Schwarzschild case. Also remarkable are the features that the
larger $J$, or the smaller $r$, the smaller the rate of
mass decrease. Accretion of phantom energy leads also to a 
decreasing of $a$ which becomes zero quickly, so that $J$ must
decrease quite more rapidly than $M$ does. For any $w$, it has been finally
seen that the rate of variation (increase for $w>-1$ and decrease for $w<-1$)
 of $J$ speeds up as one makes $r$ or $M$ larger.  

\begin{center}
\textbf{B. Generalized Chaplygin gas}
\end{center}

We shall derive now the expression for the rates $\dot{M},\dot{a},\dot{J}$ in the case of a generalized Chaplygin gas. This can be described as a perfect fluid with the equation of state\cite{chaplygin}: 
\begin{equation}\label{eq:stateChaplygin}
p=-A_{{\rm ch}}/\rho^\alpha,
\end{equation}
where $A_{{\rm ch}}$ is a positive constant and $\alpha$ is a parameter. In the particular case $\alpha=1,$ the equation of state (\ref{eq:stateChaplygin}) corresponds to a Chaplygin gas. The conservation of the energy-momentum tensor implies

\begin{equation}
\rho=\left(A_{{\rm ch}}+\frac{B}{R^{3\left(1+\alpha\right)}}\right)^{1/\left(1+\alpha\right)},
\end{equation}
with $B\equiv \left( \rho_0^{\alpha +1} -A_{{\rm ch}}\right)R_0^{3\left(\alpha+1\right)}.$ Now, from the Friedmann equation we can get

\begin{equation}
\dot{R}=\sqrt{\frac{8\pi G}{3}}R\left( t\right)\left(A_{{\rm ch}}+\frac{B}{R^{3\left(1+\alpha\right)}}\right)^{1/\left[2\left(1+\alpha\right)\right]}.
\end{equation}

Hence, 

\begin{equation}
R^{3\left(1+\alpha\right)}=\frac{B}{\left(\sqrt{\rho_0}-\sqrt{\frac{3 G}{8\pi A_{{\rm M}}^2}}I(M)\right)^{2\left(1+\alpha\right)}-A_{{\rm ch}}},
\end{equation}
\begin{equation}
R^{3\left(1+\alpha\right)}=\frac{B}{\left(\sqrt{\rho_0}-\sqrt{\frac{3 G}{2\pi^3 A_{{\rm a}}^2}}I(a)\right)^{2\left(1+\alpha\right)}-A_{{\rm ch}}},
\end{equation}
\begin{equation}
R^{3\left(1+\alpha\right)}=\frac{B}{\left(\sqrt{\rho_0}-\sqrt{\frac{6 G}{\pi}}I(J)\right)^{2\left(1+\alpha\right)}-A_{{\rm ch}}},
\end{equation}
for $M$, $a$ and $J$, respectively.
Again for the case where the dominant energy condition is
preserved, i.e. $B>0$, we obtain that $M$, $a$ and $J$ all
increase with time, $M$ and $a$ tending to constant values for moderately
large $B$. If $B$ is large enough, then whereas $M$ tends to infinity, $a$ 
approaches a larger but still finite constant value. On the other hand,
$M$ and $a$ are both seen to increase more rapidly as parameter $\alpha$ is
made smaller, with $M$ tending once again to infinity, if $\alpha$ is taken
to be sufficiently small. As to the accretion dependence on $r$ for
$B>0$, it has been checked that as $r$ is made very small, $M$ and $a$
are nearly frozen into their original values. On the contrary, for
large $r$, the evolution of $M$ will tend to match that in the Schwarzschild
case, while $a$ increases now again up to a given constant value.
If the dominant energy condition is assumed to be violated, i.e. $B<0$,
then $M$, $a$ and $J$ all decrease with time, with $M$ and $a$
always tending to minimum, nonzero constant values. Making $|B|$
larger, or $\alpha$ smaller, makes the evolution quicker and the 
final minima values for $M$ and $a$ smaller but still nonzero.
The dependence of the evolution process on $r$ in this case is quite similar
to what we already described for $B>0$, that is to say, $M$ and $a$
nearly keep their initial values for very small $r$, but both
decrease each time quicker as $r$ is increased. Also common for
$B>0$ and $B<0$ is the feature that the evolution of $M$ is damped as
we choose larger values of the angular momentum $J$.

\begin{figure}
\begin{center}
\includegraphics[width=1.0\columnwidth]{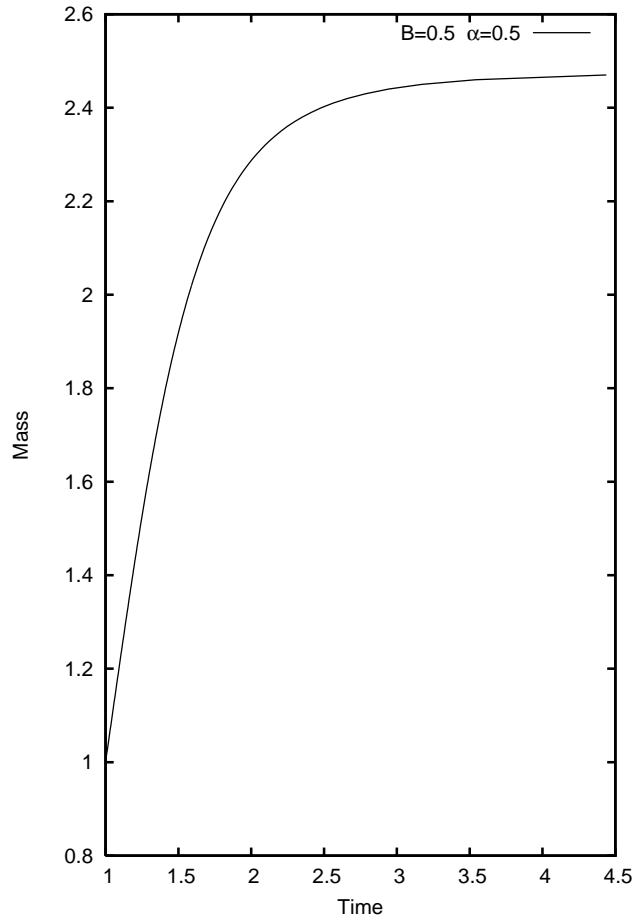}
\caption{This figure shows the behaviour of the mass of a Kerr-Newman
black hole as a function  of the cosmic time in presence of
a generalized Chaplygin gas with $B=0.5$ and $\alpha=0.5$.} \label{fig:mchaplygin}
\end{center}
\end{figure}

\begin{figure}
\begin{center}
\includegraphics[width=1.0\columnwidth]{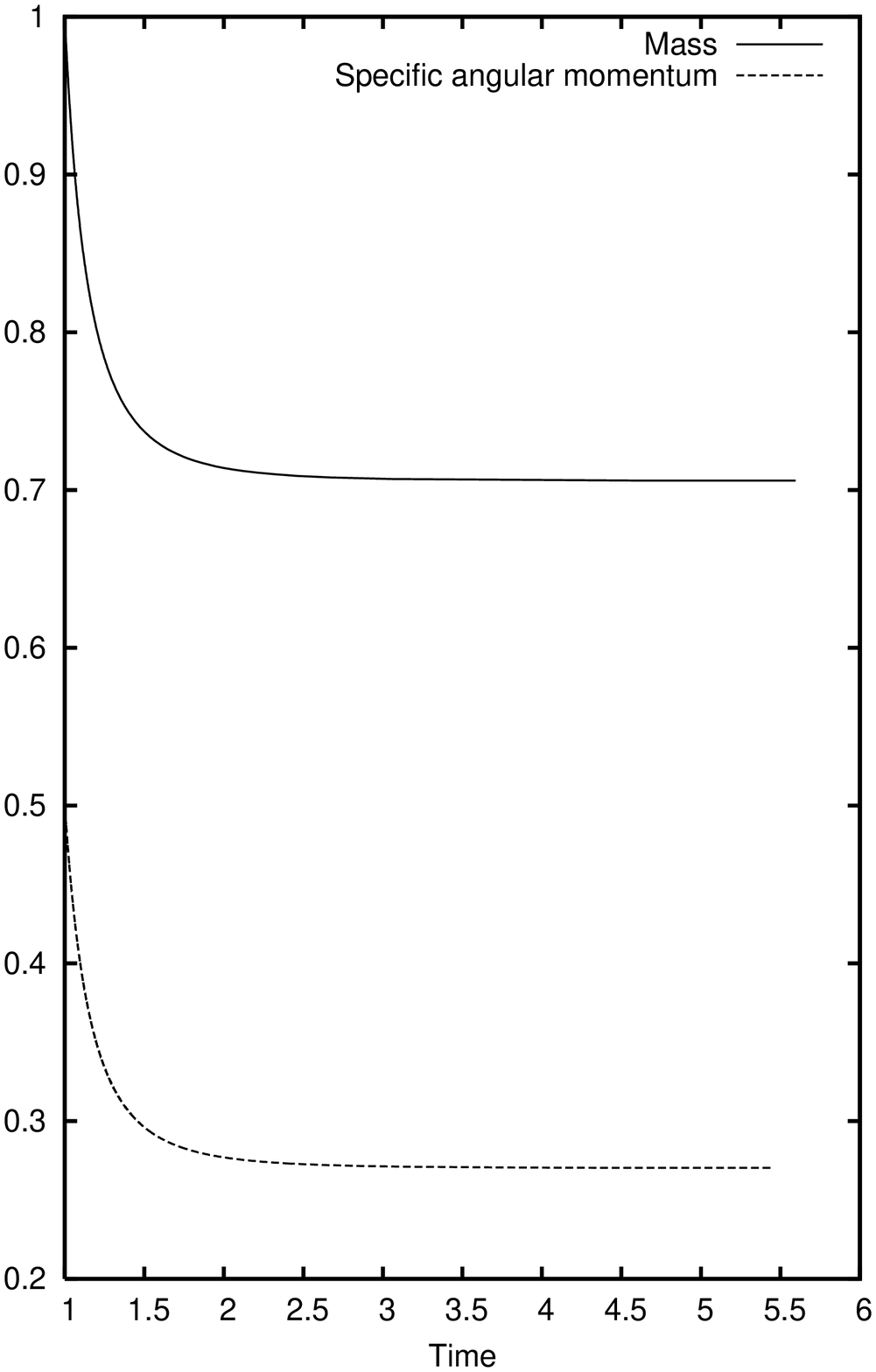}
\caption{This figure shows the behaviour of the mass and the
specific angular momentum of a Kerr-Newman
black hole as a function  of the cosmic time in presence of
a (phantom) generalized Chaplygin gas with $B=-0.5$ and $\alpha=0.5$.} \label{fig:machaplygin}
\end{center}
\end{figure}

All these behaviours have been checked by numerical calculations
which provides plots that are actually quite the same those
corresponding to the quintessence case [see Figs.~(\ref{fig:m-0.9})-(\ref{fig:j-1.1})] except (i) for the behaviour of $M$ vs time for moderate
$B>0$ and $\alpha$ far from $-1$ (in which case $M$ tends to a constant at
large $t$ [see Fig.~(\ref{fig:mchaplygin})] and (ii) for the
behaviour of $M$ and $a$ for moderate $B<0$ and $\alpha\not=-1$ (in which cases 
the studied parameters tend to nonzero constant values at large $t$ [see Fig.~(\ref{fig:machaplygin})]).

\begin{center}
\textbf{C. Super-radiance and cosmic censorship}
\end{center}

\begin{figure}
\begin{center}
\includegraphics[width=1.0\columnwidth]{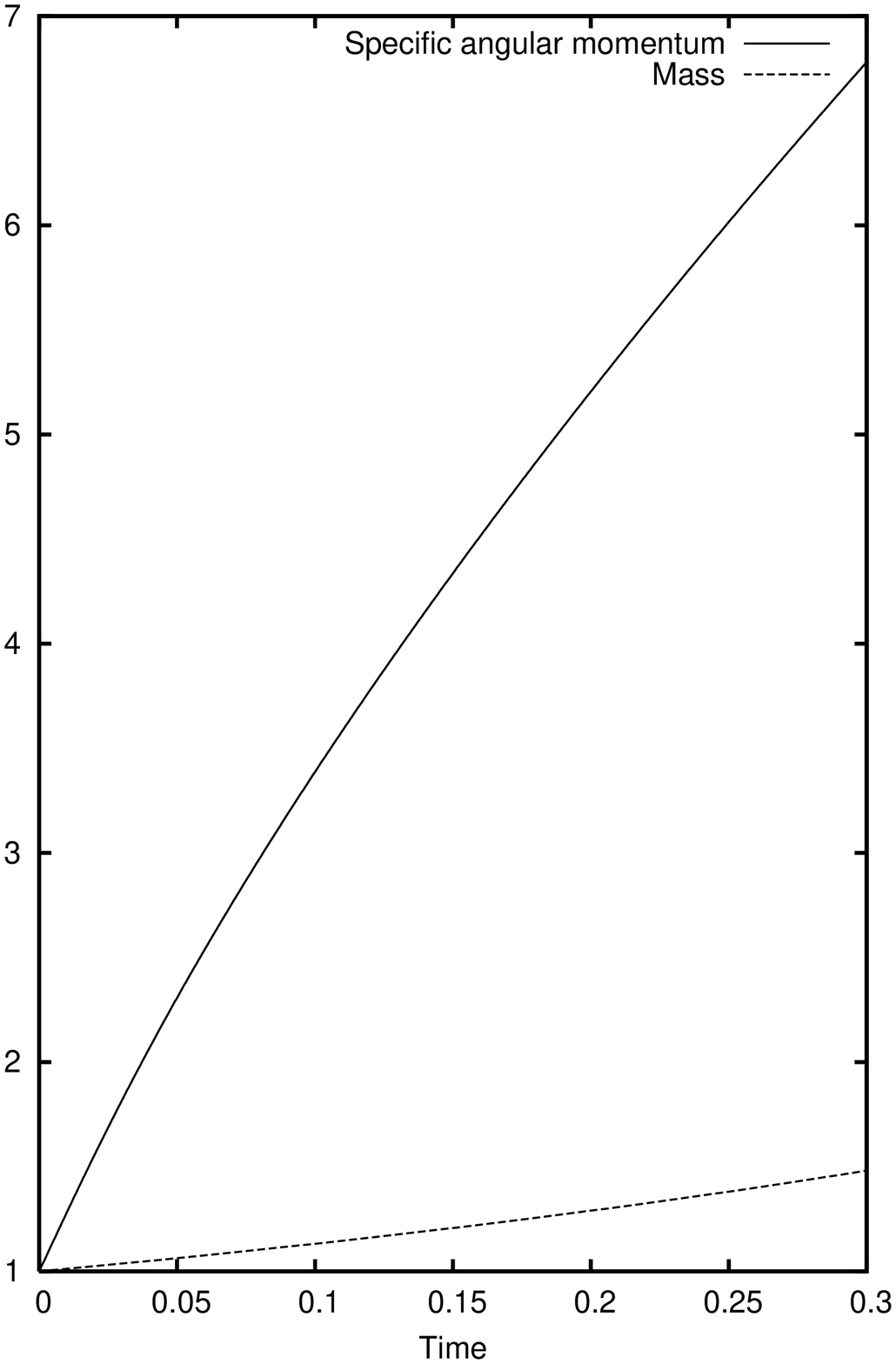}
\caption{Evolution of an extreme Kerr-Newman black hole with dark energy. This figure shows the behaviour of the mass and specific angular
momentum of a extreme Kerr-Newman
black hole as a function  of the cosmic time in the presence of
dark energy with $w=-0.9$. One can see on the figure 
that the cosmic censorship conjecture is violated.} \label{fig:extreme}
\end{center}
\end{figure}

\begin{figure}
\begin{center}
\includegraphics[width=1.0\columnwidth]{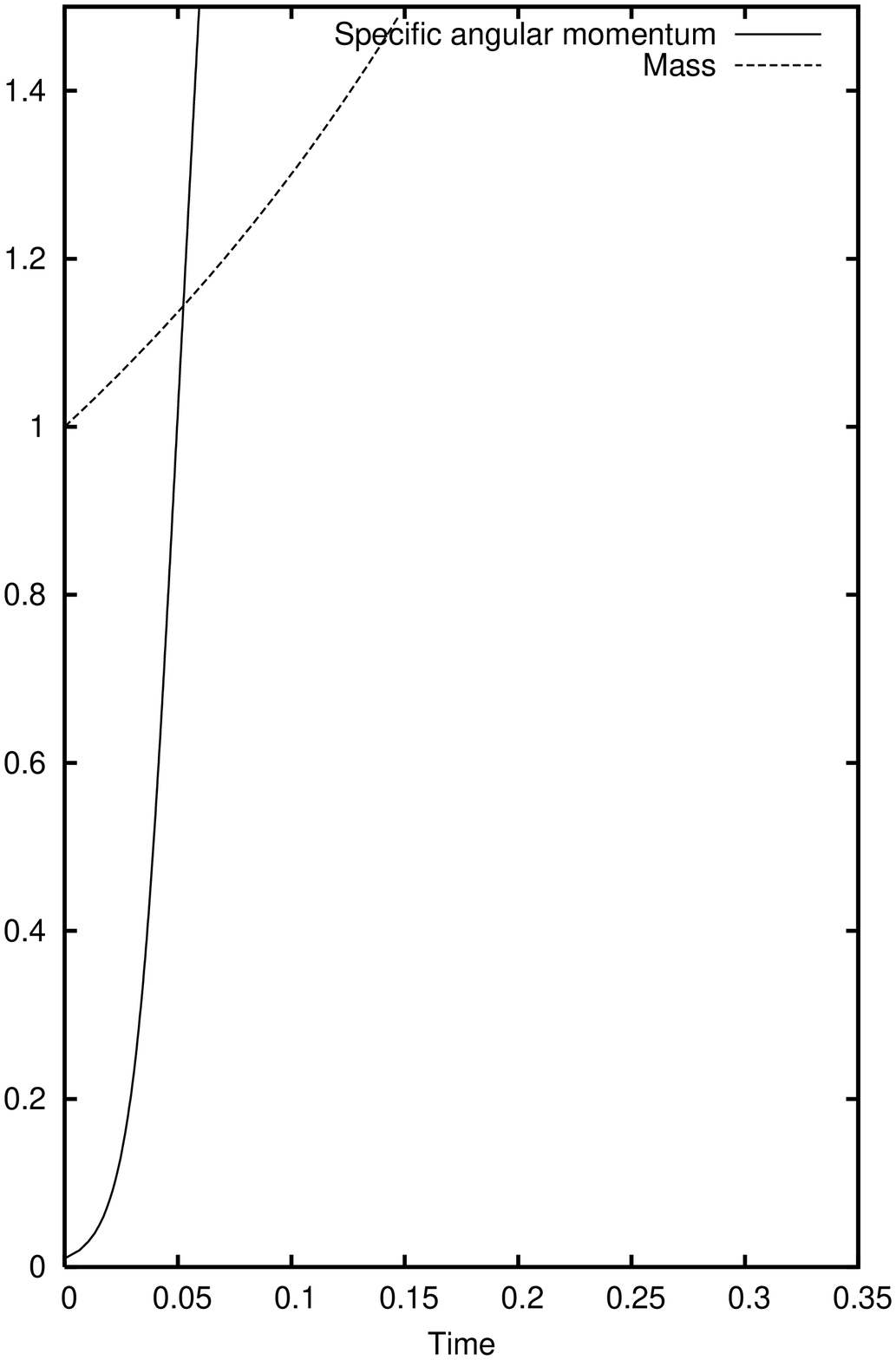}
\caption{Evolution of a Kerr-Newman black hole with dark energy. This figure shows the behaviour of the mass and specific angular
momentum of a Kerr-Newman
black hole as a function  of the cosmic time in the presence of
dark energy with $w=-0.8$. The separation between the two curves diminishes
as the initial value of $a$ is increased. One can also see on the figure 
that there exists a small 
initial time interval 
(running up to nearly $t=0.05s$ in this case) where the cosmic censorship conjecture still 
holds.} \label{fig:-0.8}
\end{center}
\end{figure}

In the case of a Kerr-Newman metric, the cosmic 
censorship conjecture\cite{Penrose} holds provided that
\begin{equation}\label{eq:conjecture}
Q^2+a^2\leq M^2.
\end{equation}
Otherwise, the Kerr-Newman black hole will show a naked singularity. It is 
interesting
to study if dark energy accretion can produce a naked singularity in this 
case. Since accretion of dark energy is a gravitatorial process, whereas
angular momentum is affected by it, electric charge is invariant under
accretion. We have pointed out above that when $P+\rho>0$, $a$ and
$M$ increase with time during accretion of dark energy. Even though we
have not a formal proof for the violation of cosmic censorship in
this case,
a numerical analysis performed for most reasonable values of $M$ and
$a$ appears to indicate that the dark energy accretion process
violates the inequality in~(\ref{eq:conjecture}) for most reasonable
situations. Actually, there always is a very small initial time interval 
where the conjecture holds, except at the extreme case where $Q^2+a^2=M^2$ [see Fig.~(\ref{fig:extreme})],
but as soon as the initial value of $a$ is taken to be significantly comparable
with that of $M$, the conjecture is almost immediately violated 
[see Fig.~(\ref{fig:-0.8})]. In the next
section we shall discuss and interpret the
reason for that violation.  

\begin{figure}
\begin{center}
\includegraphics[width=1.0\columnwidth]{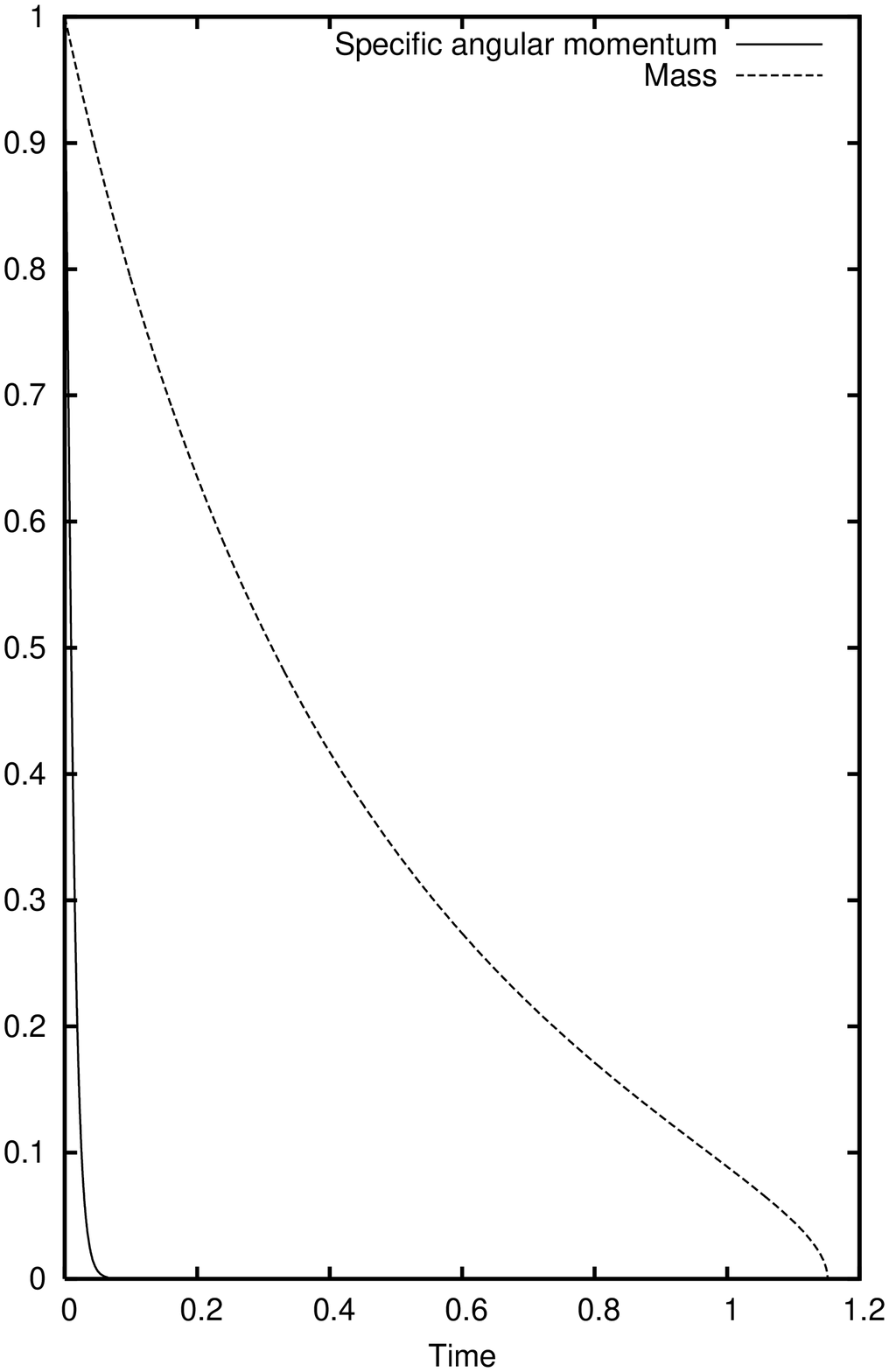}
\caption{Evolution of an extremal Kerr-Newman black hole with phantom energy. This figure shows the behaviour of the mass and specific angular
momentum of an extreme Kerr-Newman
black hole as a function  of the cosmic time in presence of
phantom energy with $w=-1.2$. The separation between the two curves increases
as the initial value of $a$ is diminished or $r$ is made larger.} \label{fig:-1.2}
\end{center}
\end{figure}

If accretion involves phantom energy, then $a$ and $M$ both decrease. In this
 case, since accretion does not affect the value of electric charge, at first sight, it could be thought that when sufficiently small values of $a$ and
$M$ are reached, Eq.~(\ref{eq:conjecture}) would no longer hold too, and cosmic censorship would be violated as well. However, it may also be expected that super-radiance of charge would act upon its value in such way that it decreased charge
during accretion of phantom energy so that Eq.~(\ref{eq:conjecture}) would
still be satisfied. Moreover, as $M$ progressively decreases the black hole
temperature should rise up and the charge super-radiance would correspondingly
speed up. Our numerical calculations appear to indicate that this is actually the case as all the  
simultaneous effects on $M$, $a$ [see Fig.~(\ref{fig:-1.2})] and $Q$ due to
dark energy accretion and Q-super-radiance seem to be 
mutually concited in such way that the cosmic censorship is preserved indeed. Obtaining an explicit, accurate expression for the relation between mass or temperature and electric charge, however is a task that contains some subtleties and therefore requires further elaboration which is left for a future consideration.
We do not consider in this paper the process of super-radiance of spin because phantom energy clearly prevails over it.

\section{An approximated accretion model}\label{approximated}

Violation of cosmic censorship in black holes dark energy accretion ($w>-1$) is a very surprising result actually, but it is perhaps not so surprising 
as  the features coming about when both rotating 
and non-rotating black holes
continue accreting such type of dark energy at sufficiently large times,
according to the accretion model used by  Babichev, Dokuchaev and Eroshenko\cite{Babichev:2004yx,Babichev:2005py} and generalized in section~\ref{sec:formalism} One of such features results e.g. from Eq.~(\ref{eq:evolutionM})
where it can be seen that the black hole mass blows up at a finite time in the
future, when the size of the universe is still finite. It follows that
the grown-up black hole will engulf the entire universe in a finite time in
the future, an implication which is rather bizarre indeed, and that it is also present when 
non-rotating black holes are considered\cite{Babichev:2004yx,Babichev:2005py}.
Nevertheless, all the predictions that have been derived for large time could
be regarded as artifacts coming from the fact that the black hole metric
used in our accretion procedure is static. Really, that procedure becomes in
such a case just an approximate description which can only be valid for a 
sufficiently short initial time interval. Therefore the results obtained in 
the present paper would just mark the tendency of the different involved parameters once the initial evolution has been overcome, but cannot be taken
for granted for large times.

Even in this case the result that cosmic censorship is violated when dark energy with $w>-1$ is being accreted cannot be justified, as that violation takes place
from the very beginning of the evolution for extreme black holes. 
According to the results displayed in Figs. (\ref{fig:m-0.9})
and  (\ref{fig:a-0.9}), there appears to be a possibility
to avoid incompatibility of a simultaneous violation of
cosmic censorship and a black hole engulfing of the universe. 
Indeed, a black hole might have the following bizarre evolution when
accretes dark energy with $w>-1$, according to these figures. It could 
violate cosmic
censorship at the beginning of its evolution and become a naked
singularity. In this stage, accretion of dark energy produces a bigger
increase of mass than specific angular momentum. Let us remember now that
specific angular momentum grows until a constant value as $t\rightarrow \infty$,
whereas the mas blow up at finite time. So, in a finite time the naked 
singularity becomes again a black hole. Next, black hole can continue its 
evolution ending in an universe engulfed by the black hole.
Thus, whereas the
evolution of black holes induced by accretion of phantom energy appears to be
quite reasonable at least on the early periods, in the case of satisfying the dominant energy condition, the accretion onto black holes seems to produce
rather unexpected results along the entire subsequent evolution.

\section{Conclusions} \label{conclusion}

In this paper we have studied the behaviour of accretion of dark energy
onto a Kerr-Newman black hole.
First, we have generalized the accretion formalism originally considered by  
 Babichev, Dokuchaev and Eroshenko\cite{Babichev:2004yx,Babichev:2005py} for the case in which the black 
hole has angular momentum and electric charge. We have applied 
such a formalism to quintessence and K-essence cosmological
fields, so as to the generalized Chaplygin gas model.  
The evolution of mass, specific angular momentum and angular momentum
when dark energy with $w>-1$ has been considered. It has been 
seen that all of these
parameters ($M$,$a$ and $J$) increase with cosmic time. The specific 
angular momentum $a$ grows up to reaching a constant value whereas $M$ 
is not bounded from above. It is also checked, in this case, that the accretion of dark energy verifying
dominant energy condition usually leads to a situation where
 the cosmic censorship is
violated. There is another feature even more surprising, i.e., the mass
of black hole blows up in a finite time and therefore black holes will
 engulf the entire universe in a finite time. These two predictions 
could be however regarded as artifacts coming from the fact that
the black hole metric used in our formalism is static. Really, 
the used  procedure could be seen just as an approximate description
which is valid only for a sufficiently short initial time interval. Therefore the
results obtained only mark the tendency of the considered parameters, and 
could well 
be not valid for large times.
Even in this case the result that cosmic censorship is violated 
when black hole accrete dark energy with $w>-1$, cannot be
justified, since that violation occurs in the very beginning of the
evolution for extreme black holes. Thus, the accretion 
of dark energy verifying $p+\rho>0$ onto black holes
seems to produce rather surprising and unexpected results.

If accretion involves phantom energy, then $a$ and $M$ both decrease
from their initial values, tending to zero as one
approaches the big rip, where the black holes 
will disappear, independently
of the initial values of their mass and angular momentum.
In  this case ($P+\rho<0$), cosmic censorship conjecture is
preserved, since super-radiance of charge and phantom energy
accretion mutually interrelated.

Whether or not the above features can be taken to imply that
phantom energy is a more consistent component than normal dark
energy with $w>-1$ is a matter that will depend on both the
intrinsic consistency of the models and the current observational 
data and those that can be expected in the future.

\section*{Acknowledgments}

\noindent We acknowledge Prof. E. Babichev for useful
explanations and correspondence. We are also grateful to V. Aldaya, A.
Ferrera, S. Robles and M. Rodr\'{\i}guez for constructive discussions and criticisms. This work was
supported by DGICYT under Research Projects BMF2002-03758 and
BFM2002-00778. 


\end{document}